\documentclass{aa}  
\pdfoutput=1
\usepackage{graphicx,url,twoopt}
\usepackage[varg]{txfonts}           
\usepackage{hyperref}                
\makeatletter
\newcommand{\bibnote}[2]{\@namedef{#1note}{#2}}
\newcommand{\biblink}[2]{\@namedef{#1link}{#2}}
\makeatother

\begin{document}
\title{H$^{12}$CN and H$^{13}$CN excitation analysis in the
  circumstellar outflow of R Scl}
\author{{M. Saberi}
     \and M. Maercker
      \and  E. De Beck
        \and W. H. T. Vlemmings
      \and H. Olofsson
      \and T. Danilovich
      }
\institute{ {Department of Earth and Space Sciences, Chalmers University of Technology, Onsala Space Observatory, 43992 Onsala, Sweden \email{maryam.saberi@chalmers.se}}
}
\date{}

\abstract 
{The $^{12}$CO/$^{13}$CO isotopologue ratio in the circumstellar envelope (CSE) of asymptotic giant branch (AGB) stars have been extensively used as the tracer of the photospheric $^{12}$C/$^{13}$C ratio. However, spatially-resolved ALMA observations of R Scl, a carbon rich AGB star, have shown that the $^{12}$CO/$^{13}$CO ratio is not consistent over the entire CSE. Hence, it can not necessarily be used as a tracer of the $^{12}$C/$^{13}$C ratio. The most likely hypothesis to explain the observed discrepancy between the $^{12}$CO/$^{13}$CO and $^{12}$C/$^{13}$C ratios is CO isotopologue selective photodissociation by UV radiation. 
Unlike the CO isotopologue ratio, the HCN isotopologue ratio is not affected by UV radiation. Therefore, HCN isotopologue ratios can be used as the tracer of the atomic C ratio in UV irradiated regions.
}
{We aim to present ALMA observations of H$^{13}$CN(4-3) and APEX\thanks{This publication is based on data acquired with the Atacama Pathfinder Experiment (APEX). APEX is a collaboration between the Max-Planck-Institut fur Radioastronomie, the European Southern Observatory, and the Onsala Space Observatory.} observations of H$^{12}$CN(2-1), H$^{13}$CN(2-1, 3-2) towards R Scl. These new data, combined with previously published observations, are used to determine abundances, ratio, and the sizes of line-emitting regions of the aforementioned HCN isotopologues.}
{We have performed a detailed non-LTE excitation analysis of circumstellar H$^{12}$CN($J$=1-0, 2-1, 3-2, 4-3) and H$^{13}$CN($J$=2-1, 3-2, 4-3) line emission around R Scl using a radiative transfer code based on the accelerated lambda iteration (ALI) method. The spatial extent of the molecular distribution for both isotopologues is constrained based on the spatially resolved H$^{13}$CN(4-3) ALMA observations.}
{We find fractional abundances of H$^{12}$CN/H$_2$ = (5.0 $\pm$ 2.0) $\times$ 10$^{-5}$  and H$^{13}$CN/H$_2$ = (1.9 $\pm$ 0.4) $\times$ 10$^{-6}$ in the inner wind ($r \leqslant$ (2.0 $\pm$ 0.25) $\times$ $10^{15}$ cm) of R Scl. The derived circumstellar isotopologue ratio of H$^{12}$CN/H$^{13}$CN  = 26.3 $\pm$ 11.9 is consistent with the photospheric ratio of $^{12}$C/ $^{13}$C $\sim$ 19 $\pm$ 6.} 
{We show that the circumstellar H$^{12}$CN/H$^{13}$CN ratio traces the photospheric $^{12}$C/$^{13}$C ratio. Hence, contrary to the$^{12}$CO/$^{13}$CO ratio, 
 the H$^{12}$CN/H$^{13}$CN ratio is not affected by UV radiation. These results support the previously proposed explanation that CO isotopologue selective-shielding is the main factor responsible for the observed discrepancy between $^{12}$C/$^{13}$C and $^{12}$CO/$^{13}$CO ratios in the inner CSE of R Scl. This indicates that UV radiation impacts on the CO isotopologue ratio. 
This study shows how important is to have high-resolution data on molecular line brightness distribution in order to perform a proper radiative transfer modelling.}


\maketitle
\section{Introduction}\label{Introduction}

During late stellar evolutionary phases, stars produce almost all heavy elements in the universe through nucleosynthesis. In the last phase of evolution of low- and intermediate-mass stars (0.8-8 $M_{\odot}$), heavy elements that are gradually built up in the inner layers are dredged up to the surface and are injected into the interstellar medium through intense stellar winds. These stars lose up to 80 percent of their  initial mass typically at rates of $10^{-8} - 10^{-4}$ $M_{\odot}$yr$^{-1}$ during the asymptotic giant branch (AGB) phase. As a result, a circumstellar envelope (CSE) of gas and dust forms around the central star, \citep[e.g.][]{Habing96, Habing03}. AGB stars can be classified by the elemental C/O ratio: C/O > 1 the carbon rich C-type stars, C/O$\sim$ 1 the S-type stars and C/O < 1 the oxygen rich M-type stars.



Molecular emission lines from CSEs are excellent probes of the physical and chemical properties of the CSE and the central star. Observations of CO rotational transitions provide the most reliable measurements of the physical parameters of the CSE such as the mass-loss rate, density structure, expansion-velocity profile, kinetic temperature, and spatial extent \citep[e.g.][]{Neri98, Schoier00, Ramstedt08, Castro-Carrizo10, DeBeck10}. Observations of other abundant molecules
set strong constraints on the chemical networks active in the CSE \citep[e.g.][]{Omont93, Bujarrabal94, Maercker08, Maercker09, DeBeck12, QuintanaLacaci16}. 


The study of isotopic ratios of evolved stars provides important information on the stellar evolutionary phases and chemical enrichment of the interstellar medium. The photospheric $^{12}$C/$^{13}$C ratio is a good tracer of the stellar nucleosynthesis.
From an observational point of view, a direct estimate of the $^{12}$C/$^{13}$C ratio is very challenging.
Hence, observing isotopologues of circumstellar carbon-bearing molecules 
are widely used to trace the elemental isotopic carbon ratio. Carbon monoxide, as the most abundant C-bearing molecule, has been extensively used to extract the carbon isotope ratio \citep[e.g.][]{Groenewegen96, Greaves97, Schoier00, Milam09},

Assuming a solitary AGB star, known processes that may affect the chemical composition of the CSE are shocks due to stellar pulsations \citep[e.g.][]{Cherchneff12} and chromospheric activity in the inner wind ($\sim r <  2.5 \times 10^{14}$ cm), gas-dust interaction in the intermediate wind ($\sim 2.5\times10^{14}<r<5 \times10^{15}$ cm), and photodissociation by the interstellar UV-radiation, associated photo-induced chemistry, and chemical fractionation processes, in the outer wind ($\sim r>5\times10^{15}$ cm), \citep[e.g.][]{Decin10}. In the case of a clumpy envelope structure, the interstellar UV-radiation can penetrate the entire envelope and affect the chemistry in the inner wind \citep[e.g.][]{Agndez10}.

The chemical processes in the inner and intermediate wind are not expected to affect the isotopologues abundance ratios. However, differences in self-shielding against UV-radiation do cause different photodissociation rates of molecules with sharp discrete absorption bands such as isotopologues of H$_2$, CO, C$_2$H$_2$, NO, \citep{Lee84}. 
Thus, isotopologue selective photodissociation by UV radiation can change the isotopologue abundance ratios in the UV irradiated regions \citep[e.g.][]{Savage96, Visser09}. Moreover, ion- molecule charge-exchange reactions in cold regions may also affect the isotopologue abundance ratios  \citep[e.g.][]{Watson76}.



Photodissociation of the molecular gas in the CSEs is thought to be dominated by the interstellar radiation field (ISRF) from the outside. ISRF is the only UV radiation field which has been considered in the modelling of AGB CSEs.
However, previous studies of UV spectra indicate the presence of a chromosphere in the outer atmosphere of carbon stars \citep{Johnson86, Eaton88}. In binary systems, active binary companions can emit UV-radiation from the inside as well \citep[{e.g.}][]{Sahai08, Ortiz16}. 
A recent search for UV emission from AGB stars has revealed that about 180 AGB stars, $\sim$ 50 $\%$ of the AGB stars observed with Galaxy Evolution Explorer (GALEX), have detectable near- or far-UV emission (Montez et al., in prep), supporting the possible existence of internal sources of UV-radiation.

Nowadays, high spatial resolution Atacama Large Millimeter/submillimeter Array (ALMA) data enable us to study the impact of UV radiation sources on the isotopologue abundance ratios in the CSEs of evolved stars more accurately.

In this paper, we derive the H$^{12}$CN/H$^{13}$CN ratio and compare it with previously reported $^{12}$CO/$^{13}$CO and $^{12}$C/$^{13}$C ratios to probe the effect of UV radiation on the CSE of R Scl.
We present the physical characteristics of R Scl in Section 2. New spectral line observations of R Scl are presented in Section 3. The excitation analysis of H$^{12}$CN and H$^{13}$CN is explained in Section 4. In Section 5, we present the results. Finally, we discuss our results and draw conclusion in Sections 6 and 7, respectively.

\section{R Sculptoris}\label{RSCL}

R Scl is a carbon-type AGB star at a distance of approximately 370 pc derived using K-band period-luminosity relationships \citep{Knapp03, Whitelock08}. It is a semi-regular variable with a pulsation period of 370 days. 
The stellar velocity $v_{LSR}^{\ast}=$ ${-19}$ km s$^{-1}$ is determined from molecular line observations.
A detached shell of gas and dust with a width of $2''$ $\pm$ $1''$ was created during a recent thermal pulse at a distance of $19.5''$ $\pm$ $0.5''$ from the central star \citep{Maercker12, Maercker14}. This shell has been extensively observed in CO and dust scattered stellar light in the optical \citep[e.g.][]{Gonzlez01, Gonzlez03, Olofsson10, Maercker12, Maercker14}. 
A spiral structure in the CSE of R Scl induced by a binary companion was revealed by $^{12}$CO($J$=3-2) ALMA observations \citep{Maercker12}. 
Moreover, a recent study of the physical properties of the detached shell by \cite{Maercker16} show that the previously assumed detached shell around R Scl is filled with gas and dust.


High spatial resolution ALMA observations of $^{12}$CO and $^{13}$CO have allowed us to separate the detached-shell emission from the extended emission of the CSE \citep{Maercker12, Vlemmings13}.
These observations reveal a discrepancy between the circumstellar $^{12}$CO/$^{13}$CO and the photospheric $^{12}$C/$^{13}$C ratios, \cite{Vlemmings13}, hereafter V13. 
They measure an intensity ratio of $^{12}$CO/$^{13}$CO$>$60 in the inner wind. Using detailed radiative transfer modelling, they show that this implies a carbon isotope ratio that is not consistent with the photospheric $^{12}$C/$^{13}$C$\sim$19 $\pm$ 6 reported by \cite{ Lambert86}. 
At the same time, they measure the intensity ratio, which varies from 1.5 to 40 in the detached shell with the average intensity ratio of $^{12}$CO/$^{13}$CO$\sim$19 which, again taking into account radiative transfer, is still consistent with the photospheric $^{12}$C/$^{13}$C ratio. 
Therefore, the circumstellar $^{12}$CO/$^{13}$CO ratio from the inner parts of the CSE, provided by the high-resolution interferometric observations, does not necessarily measure the $^{12}$C/$^{13}$C ratio. 
It has been suggested in V13 that the lack of $^{13}$CO in the recent mass loss might be due to the extra photodissociation of $^{13}$CO by internal UV radiation from the binary companion or chromospheric activity, while the more abundant $^{12}$CO would be self-shielded.

To confirm isotopologue selective photodissociation of CO as the reason for the observed discrepancy between the aforementioned CO isotopologue and C isotope ratios in R Scl, we compare the H$^{12}$CN/H$^{13}$CN and $^{12}$C/$^{13}$C ratios.
Both CO and HCN have line and continuum absorption bands in UV, respectively.
Thus, CO isotopologues are photodissociated in well-defined bands, while HCN isotopologues are photodissociated via continuum. This implies that both HCN isotopologues would be equally affected by the UV-radiation, whereas CO isotopologues have different rates of photodissociation because of  isotopologue self-shielding.
Hence, this comparison would confirm the selective photodissociation of CO as the main reason for changing the CO isotopologue abundance ratio through the CSE of R Scl.

\section{Observations}\label{observations}

\subsection{Single-dish data}

Single-dish observations of H$^{12}$CN($J$=2-1)  and H$^{13}$CN($J$=2-1, 3-2)  emission lines towards R Scl were performed using the APEX 12 m telescope, located on Llano Chajnantor in northern Chile, in July 2015. We used the SEPIA/band 5 and the SHeFI-APEX1 receivers. 
The observations were made in a beam switching mode. 
The antenna main-beam efficiency, $\eta_{\rm mb}$, the full-width half-power beam width, $\theta_{\rm mb}$, and the excitation energy of the upper transition level, $E_{up}$, at the observational frequencies for H$^{12}$CN and H$^{13}$CN are presented in Table \ref{THCN}.

The data reduction was done using XS\footnote{XS is a package developed by P. Bergman to reduce and analyse single-dish spectra. It is publicly available from ftp://yggdrasil.oso.chalmers.se}. A first-order polynomial was subtracted from the spectrum to remove the baseline. 
The measured antenna temperature was converted to the main-beam temperature using $T_{\rm mb} = T^{\ast}_{\rm A}/\eta_{\rm mb}$.


In addition to the new data presented here, we have also used previously published single-dish observations of H$^{12}$CN($J$=1-0, 3-2) made with the Swedish-ESO Submillimetre Telescope (SEST) \citep{Olofsson96} and H$^{12}$CN($J$=4-3) observed with the Heinrich Hertz Submillimeter Telescope (HHT) \citep{Bieging01}, which are summarised in Table ~\ref{THCN}.

\subsection{Interferometer data}

The ALMA observations of H$^{13}$CN($J$=4-3) were made on 14 Dec 2013, 25 Dec, 26 Apr 2014, and 24 Jul 2015 using ALMA band 7 (275-373 GHz). Figure \ref{H13CNALMA} shows the H$^{13}$CN($J$=4-3) integrated flux density over the velocity channels, a zero-moment map. 
The primary flux calibration was done using Uranus and bootstrapped to the gain calibrator J0143-3206 (0.27 Jy beam$^{-1}$) and J0106-4034 (0.23 Jy beam$^{-1}$). Based on the calibrator fluxes, the absolute flux has an uncertainty of around 10 $\%$. 
The data reduction was done with the Common Astronomy Software Application (CASA). More details of the data reduction will come in Maercker et al. in prep.
The tasks "imsmooth" and "immoments" in CASA were used to smooth the image to $0.13'' \times 0.13''$ resolution and to integrate over velocity channels, respectively.

\begin{figure}[t]
  \centering
  \includegraphics[width=80mm]{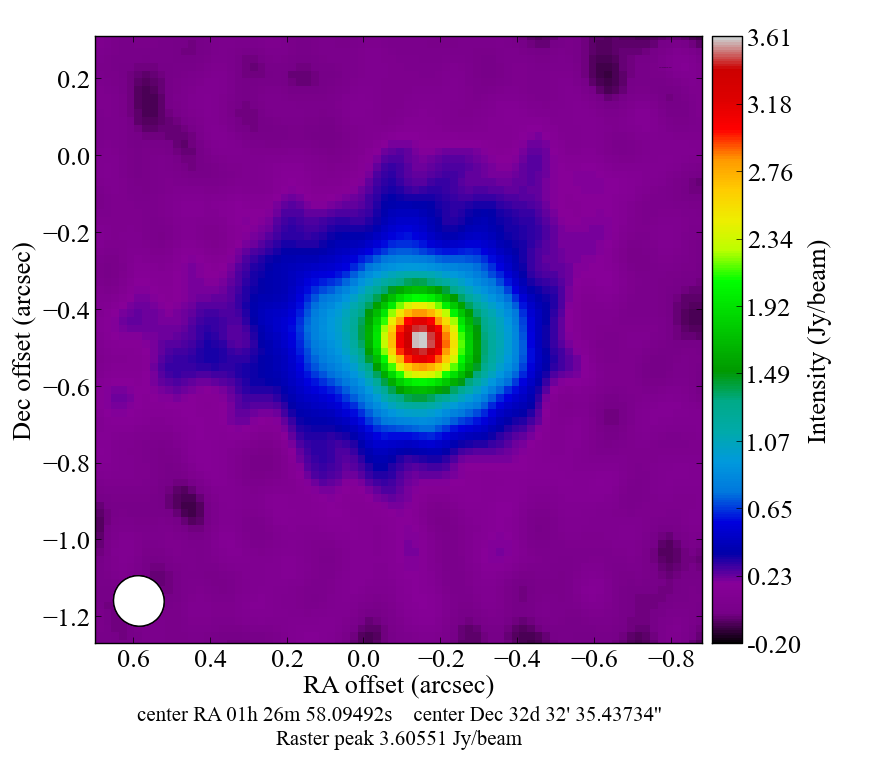}
  \caption[]{\label{H13CNALMA}
     Zero-moment H$^{13}$CN(4-3) map of R Scl observed with ALMA. The ALMA beam size is shown in the bottom left corner.}
\label{H13CNALMA}  
\end{figure}

\begin{table}[t]
  \centering
  \setlength{\tabcolsep}{5.5pt}
    \caption{Observations of HCN towards the circumstellar envelope of R scl.}
  \begin{tabular}{@{} ccccccc@{}}
\hline
Trans. & Tel. & Freq. [ GHz]& $\eta_{\rm mb}$ & $\theta_{\rm mb}['']$ & $E_{\rm up}$ [K] & Ref \\
 \hline
H$^{13}$CN\\
\hline
$J$=2-1 & APEX & 172.6 & 0.80 & 35 & 12.43 & 3
 \\ 
$J$=3-2 & APEX & 259.01 & 0.75 & 30 & 24.86 & 3
 \\
$J$=4-3 & ALMA & 354.3 &  &  & 41.43 & 3
 \\
 \hline
H$^{12}$CN\\
\hline
$J$=1-0 & SEST & 88.6 & 0.75 & 55 & 4.25 & 1
 \\   
$J$=2-1 & APEX & 177.2 & 0.80 & 35 & 12.76 & 3
 \\ 
$J$=3-2 & SEST & 265.9 & 0.50 & 21 & 25.52 & 1
 \\
$J$=4-3 & HHT & 354.5 & 0.48  & 22 & 42.53 & 2
 \\
   \hline
 \end{tabular}
\tablefoot{References: 1. \cite{Olofsson96}, 2. \cite{Bieging01} and 3. This work}
 \label{THCN}
\end{table}

\section{Excitation analysis}\label{Excitation Analysis}

\subsection{Spectroscopic treatment of HCN}\label{Spectroscopic}


The molecule HCN is linear and polyatomic, with three vibrational modes: the H-C stretching mode $\nu_1$ at 3 $\mu$m, the bending mode $\nu_2$ at 14 $\mu$m, and the C-N stretching mode $\nu_3$ at 5 $\mu$m. In our modelling, we take into account the $\nu_1$=1 and the $\nu_2$=1 states, while we neglect the $\nu_3$ mode since this includes transitions that are about 300 times weaker than $\nu_1$=1 \citep{Bieging84}.
The nuclear spin (or electric quadrupole moment) of the nitrogen nucleus leads to a splitting of the rotational levels into three hyperfine components. Furthermore, a splitting of the bending mode occurs due to rotation when the molecule is bending and rotating simultaneously. This is referred to as $l$-type doubling. This interaction between the rotational and bending angular momenta results in $l$-doubling of the bending mode into two levels $01^{1c}0$ and $01^{1d}0$. 

In our modelling, the excitation analysis includes 126 energy levels. Hyperfine splitting of the rotational levels for $J$=1 levels are included. In each of the vibrational levels, rotational levels up to $J$=29 are considered, and the $l$-type doubling for $\nu_2$=1 transitions are also included. 
For excitation analysis, the same treatment as \cite{Schoier11, Danilovich14} with small adjustments was used.

The HCN vibrational states are mainly radiatively excited \citep[e.g.][]{Lindqvist00, Schoier13}. In our modelling, the radiation arises from the central star, which is assumed to be a black body, and from the dust grains, which are distributed through the CSE.

\subsection{Radiative transfer model}\label{RT}

To determine the HCN isotopologue abundances and constrain their distributions in the CSE, a non-local thermodynamic equilibrium (LTE) radiative transfer code based on the accelerated lambda iteration (ALI) method was used. The ALI method is described in detail by \cite{Rybicki91}. The code has been implemented by \cite{Maercker08} and was previously used by \cite{Maercker09, Danilovich14}.

The CSE around R Scl is assumed to be spherically symmetric and is formed due to a constant mass loss rate $\dot{M}\sim 2\times10^{-7} $ $M_{\odot}$ yr$^{-1}$ \citep[][hereafter W04]{Wong04}. The inner radius is assumed to be located at $r_{\rm in} = 10^{14}$ cm. The H$_2$ number density is calculated assuming a constant mass loss rate, as in \cite{Schoier01}.


The gas-expansion profiles is assumed to be:
\begin{equation}
v_{\rm exp}\left(r\right) = v_{\rm min} + \left( v_\infty - v_{\rm min} \right) \left( 1 - \frac{r_{\rm in}}{r} \right)^{\rm b},
\label{V_exp}
\end{equation}
where $v_\infty$ is the terminal expansion velocity and $v_{\rm min}$ is the minimum velocity at $r_{\rm in}$, which is taken as the sound speed 3 km s$^{-1}$ and b determines the shape of the radial velocity profile. 
A detailed discussion on the determination of the two free parameters $v_\infty$ and b based on the shape of line profiles at radial offset positions from the ALMA observations is presented in Sect. \ref{HCN-V}.




The radial distribution of the dust temperature is derived based on SED modelling (Maercker, private communication) to be a power-law given by:
\begin{equation}
T(r) = T_0  \left(\frac{r_0}{r}\right)^{0.38},
\end{equation}
where $T_0$ = 1500 K  and $r_0=7\times 10^{13}$ cm are the dust condensation temperature and radius. 
Since ALI does not solve the energy balance equation, the same temperature profile as the dust temperature was used to describe the kinetic temperature of the gas. We also ran models using the gas temperature profile derived by W04, which led to no significant change in the results.

The HCN fractional molecular abundance relative to H$_2$ $(n_{\rm HCN}/n_{\rm H2})$ is assumed to have a gaussian distribution:
\begin{equation}
f(r) = f_0  exp\left(-\left(\frac{r}{R_e}\right)^2\right),
\end{equation}
where $f_0$ denotes the initial fractional abundance and $R_e$ is the $e$-folding radius for HCN, the radius at which the abundance has dropped to 1/$e$ (37\%). 
The stellar parameters are presented in Table \ref{Tpara}.

\begin{table}[t]
  \caption{The stellar parameters that are used in the radiative transfer modelling of the H$^{12}$CN and H$^{13}$CN isotopologues around R Scl.}
  \begin{tabular}{lll }
\hline
Fixed model parameters & & Ref  \\
 \hline
Distance (pc) & 370 & 1
 \\   
Mass-loss ($M_{\odot}$ yr$^{-1}$)  & $2 \times 10^{-7}$ &  2
 \\ 
 Luminosity ($L_{\odot}$) & 6800 & 3
 \\
 Turbulent velocity (km/s) & 1 & 3
 \\
$T_{\rm star}$ (K) & 2500 & 3
\\
$r_{\rm in}$ (cm)& $10^{14}$ & 3
\\
$r_{\rm out}$ (cm) & 3 $\times$ $R_e$ & 3
\\
   \hline
 \end{tabular}
  \tablefoot{References: 1. \cite{Knapp03, Whitelock08} , 2. \cite{Wong04} and 3. This work}
  \label{Tpara}
\end{table}

\section{Results}\label{Results}

\subsection{Radial expansion velocity profile}\label{HCN-V}

\begin{figure}[t]
  \centering
   \begin{minipage}[b]{0.41\textwidth}
    \includegraphics[width=\textwidth]{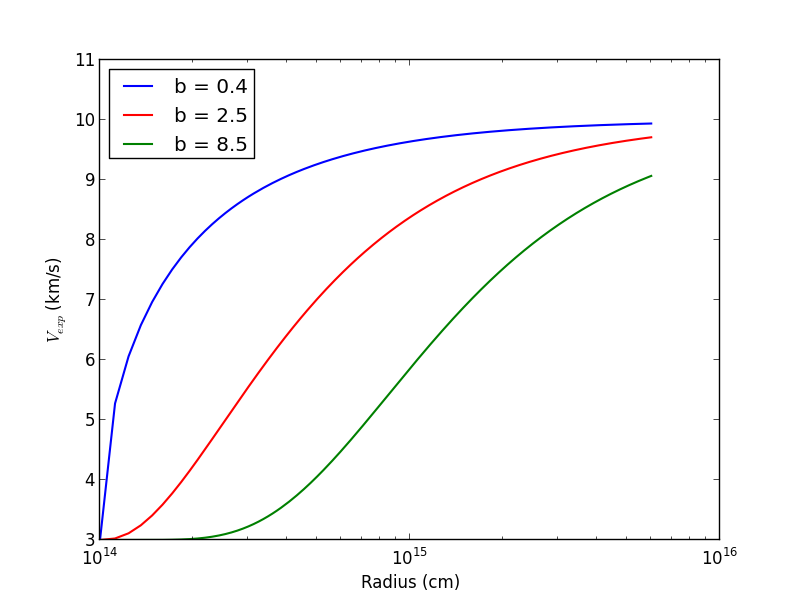}
     \caption{Gas radial expansion velocity profiles for the $v_\infty$=10 km s$^{-1}$ and b values of 0.4, 8.5 and 2.5. The model with b = 2.5 leads to the best fits to the line shape of spectra at all radial offset positions from the centre of the star.}
      \label{Vexp2}
  \end{minipage}
  \hfill
  \begin{minipage}[b]{0.41\textwidth}
    \includegraphics[width=\textwidth]{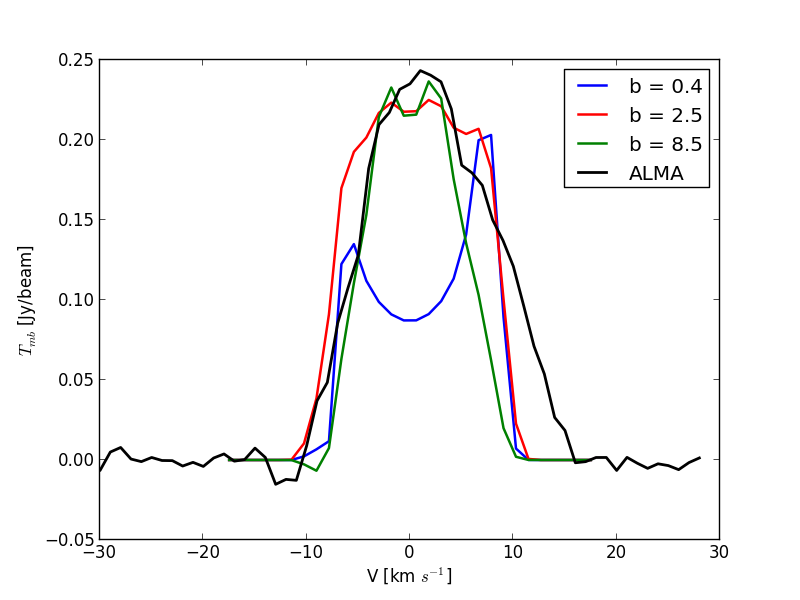}
    \caption{Comparison of the H$^{13}$CN(4-3) intensity profile at the centre of the star from ALMA observations and three models with b = 0.4, 2.5 and 8.5. The model with b = 2.5 reasonably fits the line spectra, while models with b = 8.5 and 0.4 are not consistent with the observations at the centre of the star.
    }    
         \label{Vexp1}
  \end{minipage}
\end{figure}

We use the spatially resolved ALMA observations of H$^{13}$CN(4-3) to constrain two free parameters, b and $v_\infty$,  in the gas radial expansion velocity profile, Eq.\ref{V_exp}. 
We extracted intensities at a series of offset positions sampling every independent beam from the ALMA observations and the corresponding intensities from the modelling results.  
A series of models with b and $v_\infty$ changing in the ranges $0.4 < b < 8.5$ and $8.5 < v_\infty < 13$ were run to get good fits to the line shape of spectra at all positions. The model with b = 2.5 and $v_\infty$=10 km s$^{-1}$ leads to the best fits to the line shape of the H$^{13}$CN(4-3) spectra at the offset positions from the centre. 
The velocity profiles with b = 0.4, 2.5 and 8.5 values are shown in Fig. \ref{Vexp2}. 
Models with b < 2.5 did not reproduce the shape of H$^{13}$CN(4-3) line profiles in the inner part of the envelope (r $\leqslant $0.2$ '' $); they predicted double-peaked line profiles contrary to the observed spectra. On the other hand, models with large b values reproduce narrow line shapes, which are also not consistent with the observations.  
To illustrate this, we plot the H$^{13}$CN(4-3) intensity profiles at the centre of the star from ALMA observations and from three models with b = 0.4, 2.5 and 8.5 in Fig. \ref{Vexp1}. 
There is red-shifted excess emission in the ALMA H$^{13}$CN(4-3) that can not be reproduced by our spherically symmetric model.
The models with b = 0.4 and 8.5 can predict the total intensity of the H$^{13}$CN(4-3) pretty well, but they fail to reproduce the line shapes at the inner part of the envelope (r$\leqslant$0.2$ '' $).
Without having spatially resolved observations, it is not possible to find out such effects and precisely constrain the gas expansion velocity profile.
 

It should be noted that the wind-expansion velocity of R Scl likely changes over time \citep{Maercker16}. The gas expansion velocity profile used here hence mostly describes the change in terminal expansion velocity, rather than the acceleration due to radiation pressure on dust grains. Typical values for the exponent in accelerated-wind profiles are approx. 1.5 for M-type stars \citep[e.g.][]{Maercker-2-16} and even less for carbon stars. The fact that we derive a larger exponent confirms that the wind-expansion velocity of R Scl has been declining since the formation of the shell.


\subsection{Radial distributions of HCN isotopologues}\label{HCN-R}

The molecular photodissociation by UV-radiation is the dominant process controlling HCN survival throughout the CSE. Consequently, it is the most important factor in determining the size of the HCN molecular envelope in the CSE. Since both HCN isotopologues are equally affected by UV radiation, the same molecular distribution is expected for both H$^{12}$CN and H$^{13}$CN isotopologues.

To estimate $R_e$, we use the spatially resolved ALMA observations of H$^{13}$CN(4-3) which strongly constrains the $e$-folding radius. 
We ran 102 models with the fixed parameters detailed in Table~\ref{Tpara} and simultaneously varying $f_0$ and $R_e$ over the ranges $6\times10^{-7} < f_0 < 3 \times10^{-6}$ and $1\times10^{15} < R_e < 6\times10^{15}$ cm (6 values for $f_0$ and 17 values for $R_e$).
To select the model with the best $R_e$, a reduced $\chi^2$ statistic is used. The reduced chi-square is defined as $\chi^2_{red} = \chi^2 / (N-2)$, where 2 is the number of free parameters in our modelling. 
In $\chi^2_{red}$ calculation, we only use the average intensities at the radial offset positions (9 positions) from four directions from the centre of the star which are shown in Fig.~\ref{Radi}.
The model with the minimum $\chi^2_{red}$ has the $e$-folding radius $R_e = (2.0 \pm 0.25) \times 10^{15}$ cm. The cited error is for 1$\sigma$ uncertainty derived from the $\chi^2_{red}$ distribution.

\begin{figure}[t]
  \centering
  \includegraphics[width=78mm]{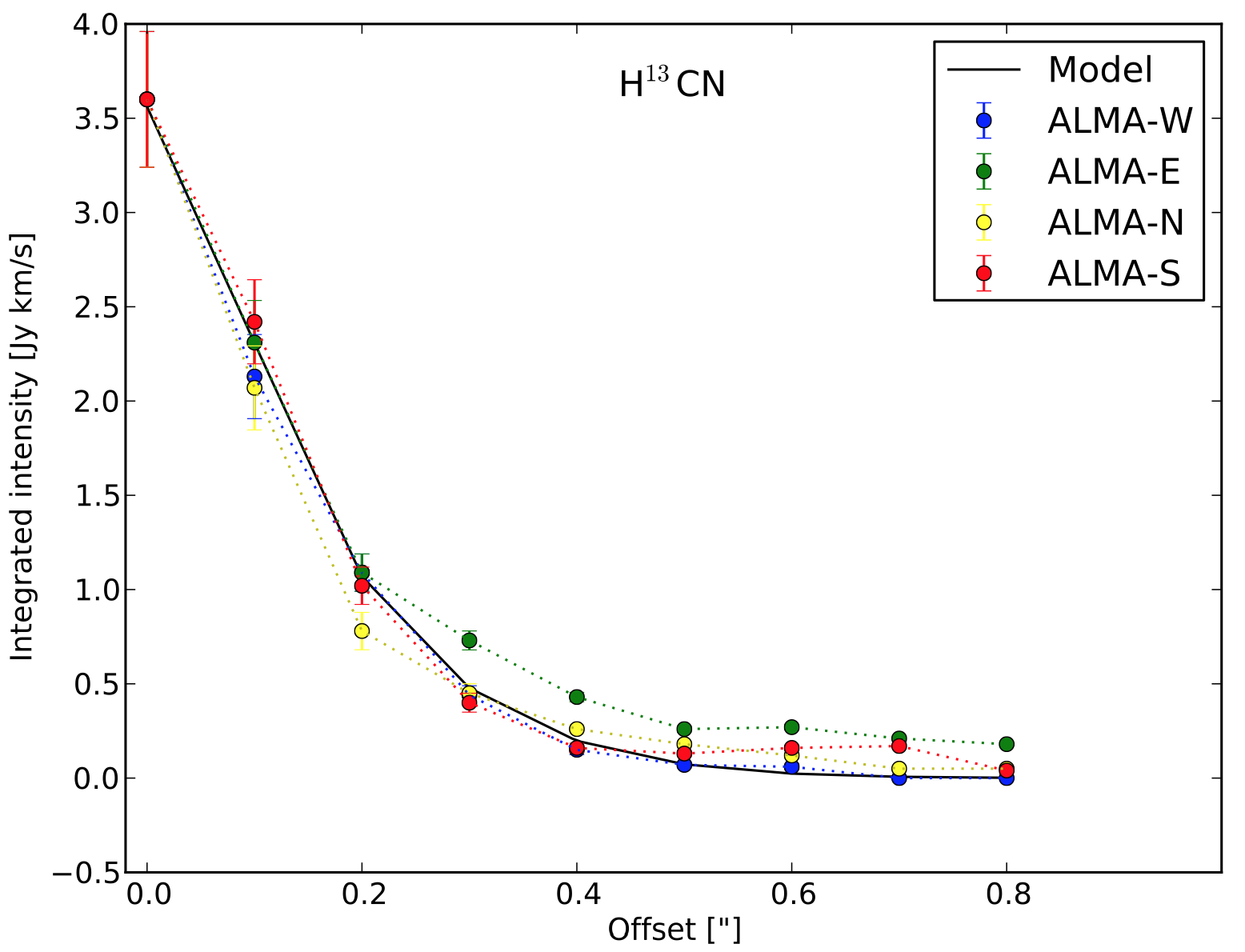}
  \caption[]{\label{Radi}
    Comparison of the integrated ALMA intensities of H$^{13}$CN(4-3) at radial offset points in the CSE of R Scl towards the west, east, north and south from the centre of the star with the best-fitting model which constrains the HCN molecular distributions at $R_e = 2 \times 10^{15} cm$. Error-bars on the observational points show 10$\%$ uncertainty on the flux calibration.}
\label{Radi}  
\end{figure}

\begin{figure}[t]
  \centering
  \includegraphics[width=80mm]{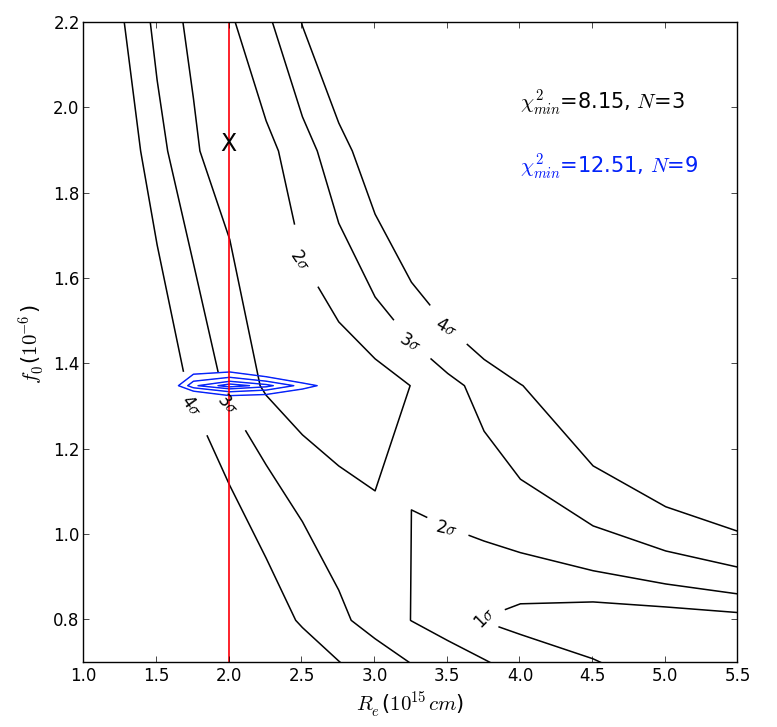}
  \caption[]{\label{chi2map}
    Two $\chi^2_{red}$ maps calculated using the average intensities at radial offset points from H$^{13}$CN(4-3) ALMA observations (blue), and the total intensities of H$^{13}$CN(2-1, 3-2,4-3) (black). The min $\chi^2_{red}$ values and the number of observational constraints that are used in the calculations are written. Contours are plotted at the 1,2,3 and 4 $\sigma$ standard deviation levels from the minimum $\chi^2_{red}$. The best-fitting model for H$^{13}$CN is shown by a black X.
  }
\label{chi2map}  
\end{figure}

\subsection{The H$^{13}$CN abundance}\label{H13CN}



\begin{figure}[t]
  \centering
   \begin{minipage}[b]{0.51\textwidth}
    \includegraphics[width=\textwidth]{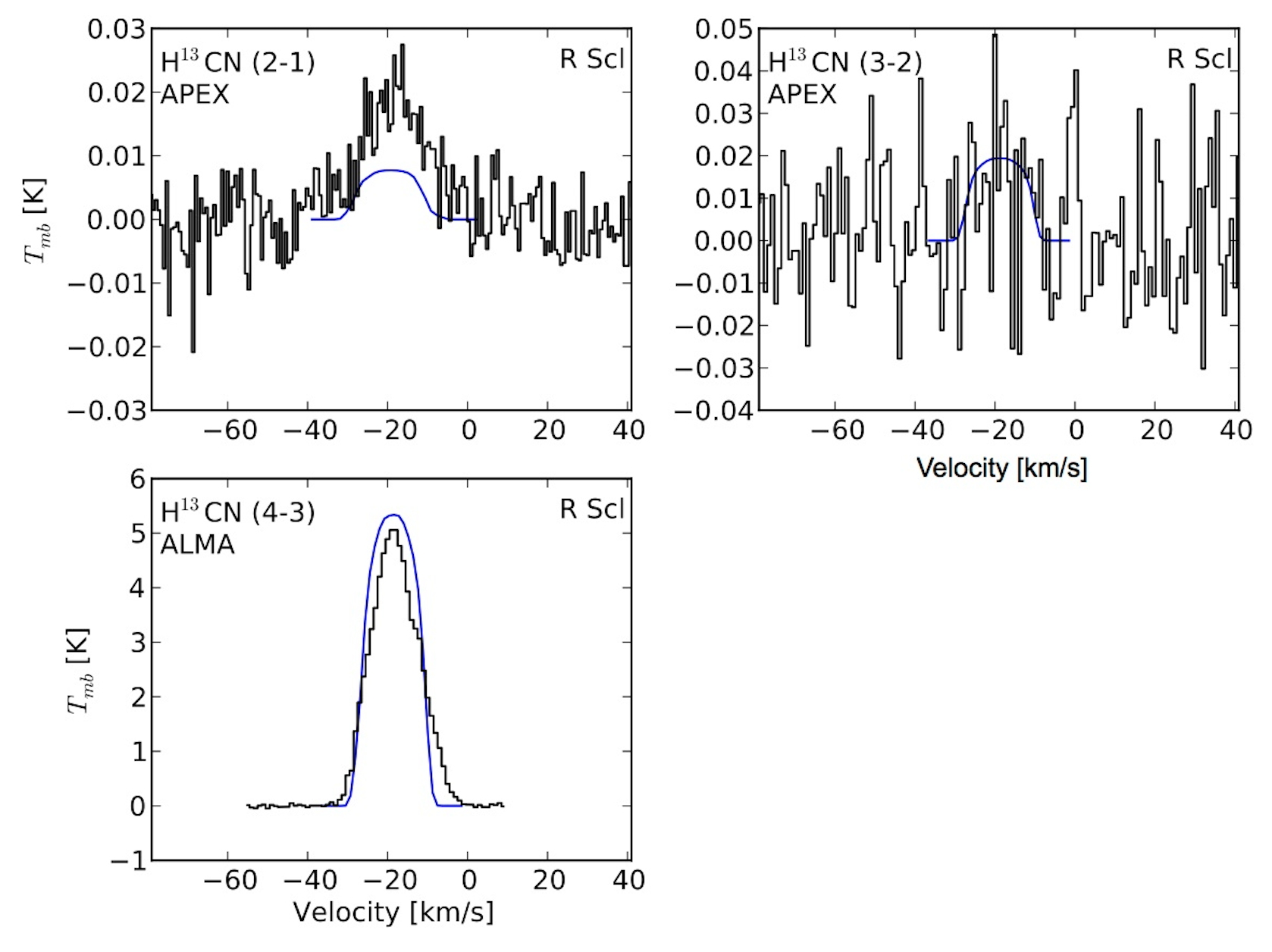}
     \caption{Line emission of H$^{13}$CN towards R Scl (black) overlaid with the best-fitting model (blue) which has values of $f_0 = 1.9 \times 10^{-6}$ and $R_e = 2.0 \times 10^{15}$ cm. Molecular transitions and the telescope used to get data are written in each panel. The H$^{13}$CN(4-3) ALMA spectrum is extracted with $2''$ beam size.}
      \label{h13cnSpectra}
  \end{minipage}
  \hfill
  \begin{minipage}[b]{0.47\textwidth}
    \includegraphics[width=\textwidth]{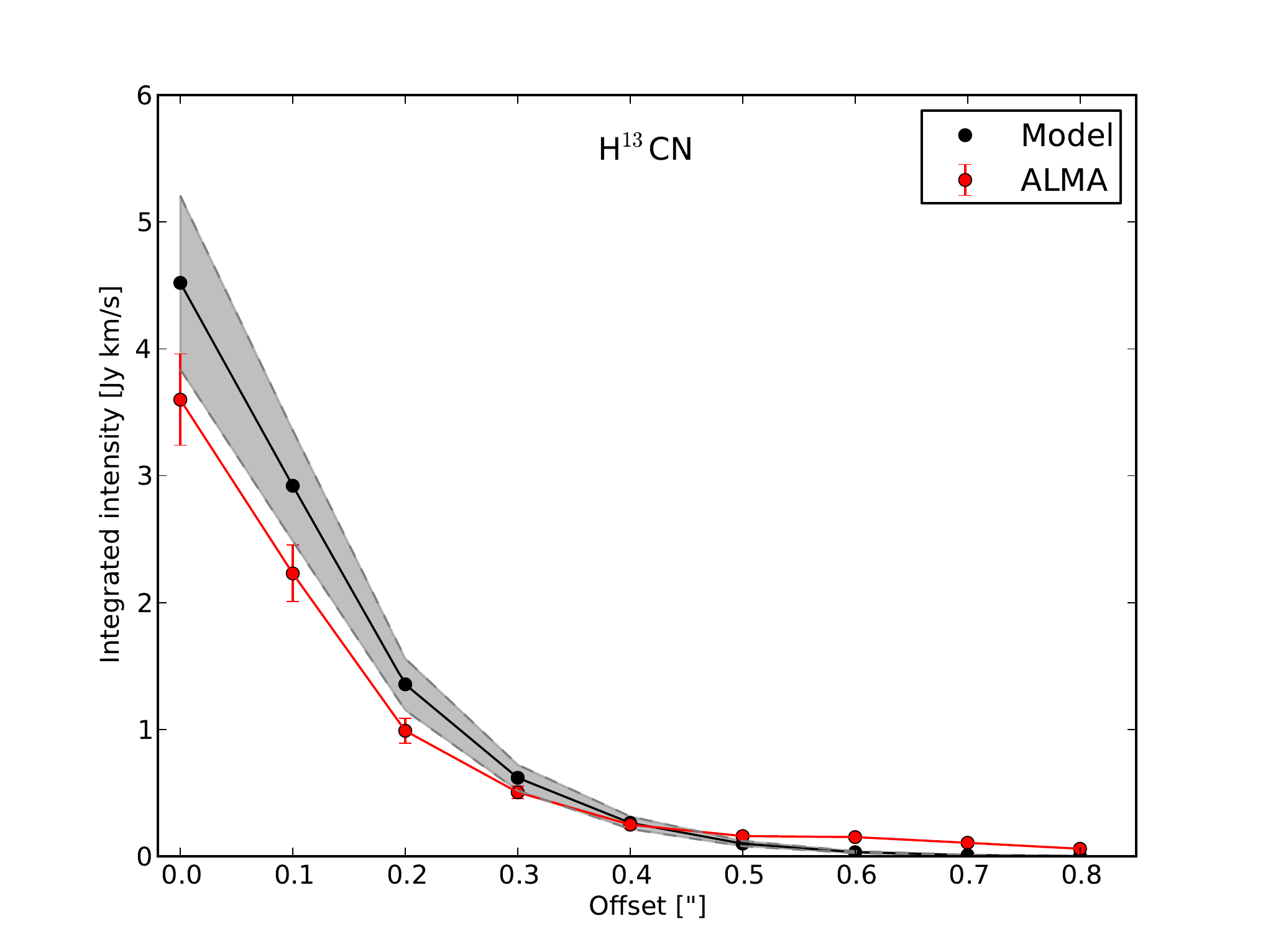}
    \caption{Comparison between the average of integrated intensities of the H$^{13}$CN(4-3) at radial offset positions with best-fitting model which has values of $f_0 = 1.9 \times 10^{-6}$ and $R_e = 2.0 \times 10^{15}$ cm. The grey region shows 1 $\sigma$ confidence level for $R_e$. }    
         \label{h13cnRadial}
  \end{minipage}
\end{figure}

To estimate the H$^{13}$CN initial value $f_0$, we use the total intensities of all three observed lines H$^{13}$CN(2-1, 3-2, 4-3) in $\chi^2$ calculation.
The uncertainty of the observed lines in $\chi^2$ calculation, $\sigma$, is assumed to be 20$\%$ for H$^{13}$CN(2-1) APEX data, 10$\%$ for H$^{13}$CN(4-3) ALMA data and 100$\%$ for H$^{13}$CN(3-2) undetected line. The H$^{13}$CN(3-2) undetected line was only used to put a limit on the adjustable parameters. Since various rotational transitions come from different regions of the envelope, changing the parameters in the model has a different effect on these transitions.
The $\chi^2$ map is shown in Fig.~\ref{chi2map}, accompanied with the $\chi^2_{red}$ map derived using H$^{13}$CN(4-3) observations to constrain the $R_e$. 
The best-fitting model is chosen among the six models with $R_e = 2.0 \times 10^{15}$ cm and $6\times10^{-7} < f_0 < 3 \times10^{-6}$, which are shown with a red line in the map. 
The best model with the minimum $\chi^2$ has the fractional abundance $f_0 = (1.9 \pm 0.5) \times 10^{-6}$. 
The H$^{13}$CN spectra accompanied with the best-fitting model are presented in Fig. \ref{h13cnSpectra}. We also compare the average intensities at the radial offset positions from ALMA observations and the best-fitting model results in Fig. \ref{h13cnRadial}.

Our excitation analysis shows differences between the kinetic and the excitation temperatures, meaning that all observed lines are formed under non-LTE conditions. The maximum tangential optical depth varies from $\sim$ 0.09 up to 0.7 for the $J=2-1$ to $J=4-3$ lines, indicating that all lines are optically thin.

\subsection{The H$^{12}$CN abundance}\label{HCN}

To estimate the H$^{12}$CN fractional abundance $f_0$, we ran 11 models with $R_e = 2.0 \times 10^{15}$ cm and  $f_0$ varying in the range of $1\times 10^{-5} <f_0< 6.5\times 10^{-5}$.
Figure \ref{h12cnSpectra} shows the H$^{12}$CN spectra overlaid with the best-fitting model with the minimum $\chi^2_{red} = 0.07$. This model has a fractional abundance of $f_0 = (5.0 \pm 2)\times 10^{-5}$. 
In the $\chi^2$ calculation, we consider the integrated intensity of the H$^{12}$CN(2-1, 3-2, 4-3) lines. Since there is evidence of maser emission in the H$^{12}$CN($J$=1-0) line \citep[e.g.][]{Olofsson93, Olofsson98, Lindqvist00, DinhVTrung00, Shinnaga09, Schoier13}, we do not include in the $\chi^2$ statistic.

The difference between the kinetic and the excitation temperatures for H$^{12}$CN indicates that all observed lines are formed under non-LTE conditions. The maximum tangential optical depth varies from $\sim$ 8 up to 30 for $J=1-0$ to $J=4-3$ lines, indicating that all lines are optically thick. 

\begin{figure}[t]
  \centering
  \includegraphics[width=89mm]{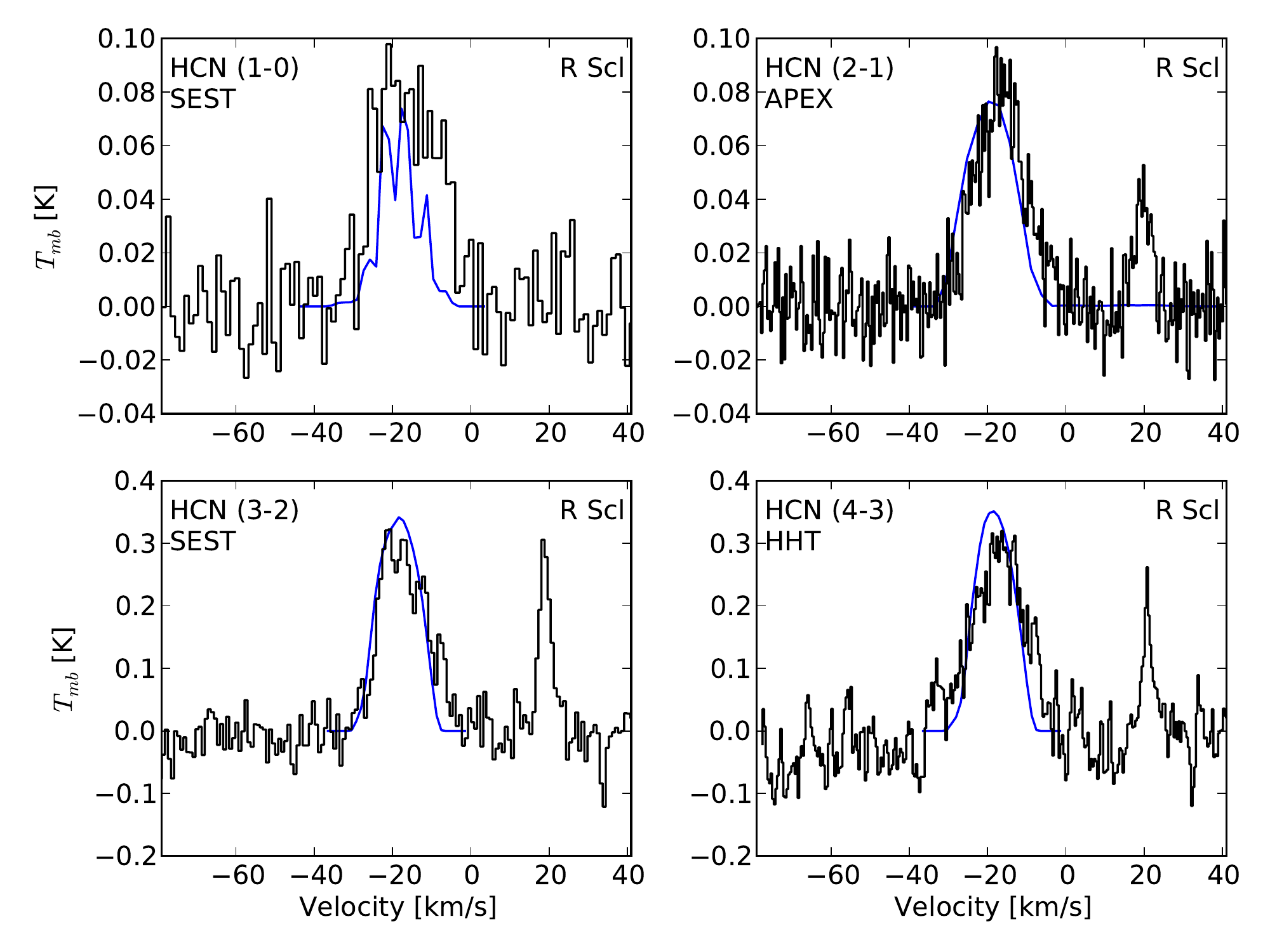}
  \caption[]{\label{h12cnSpectra}
    Line emission of H$^{12}$CN towards R Scl (black) overlaid with the best-fitting model (blue). Molecular transitions and the telescope used to get data are written in each panel. Second peaks in $J=$2-1, 3-2, 4-3 transitions are due to maser emission in the (0$1^{1c}$0) vibrational state.}
\label{h12cnSpectra}  
\end{figure}

\section{Discussion}\label{Discussion}

\subsection{Comparison of the best-fitting criteria based on single-dish and interferometric observations}

Comparison between two $\chi^2$ maps in Fig. \ref{chi2map} shows that the best-fitting criteria based on the high-resolution interferometric and low-resolution single-dish observations are different.
The single-dish observations require a larger $e$-folding radius with less fractional abundance, while interferometric observations require a more compact envelope with higher abundance. This illustrates the importance of spatially resolved images of even a subset of the transitions used in molecular modelling. For comparison, a model selected only on the single-dish spectra is presented in the Appendix.
 
The H$^{12}$CN modelling for R Scl by W04 also implements the different best-fitting criteria based on the single-dish and interferometric observations. They
require a more compact envelope with less abundance to fit the spatially-resolved H$^{12}$CN(1-0) ATCA observations, while they can not fit the single-dish spectra using the same condition (see figures 7 and 8 in W04).

A possible explanation for this discrepancy might be the spherically symmetric molecular distribution considered in our models, especially considering R Scl is known to have a binary companion whose influence gives rise to the spiral pattern observed in CO \citep{Maercker12}. Constraining this asymmetry in the inner wind is, with the currently available observations, not yet possible. However, while the asymmetry could affect the derived abundances, it does not affect our determination of the isotopologue ratio, which is the main aim of this work.

\subsection{Asymmetry in the CSE }


Our modelling is based on assuming a spherically symmetric envelope around R Scl. This is a first order approximation, and the east-west asymmetry in the CSE as seen in Fig. \ref{H13CNALMA} possibly due to the binary companion is not taken into account.
Indeed, we are not able to precisely constrain the additional free parameters required to fully describe the physical condition such as the molecular distribution profiles from the available data.   
To find out the effect of this asymmetry in determining the adjustable parameters, we calculate the $\chi^2_{red}$ statistic in four directions from the central star.
As seen in Fig. \ref{chi2map2},  $\chi^2_{red}$ in west, north and south lead to approximately the same range of adjustable parameters, while the contour map from the east shows a larger $R_e$, which is expected to be due to the elongation of the molecular gas towards the east as seen also in Fig. \ref{H13CNALMA}.

The east-west elongation of the molecular distribution could also be an explanation of the underestimate of the lower-$J$ HCN emission. Assuming that the lower-$J$ transitions are predominantly excited in the more extended aspherical region, a spherical symmetric model will either overestimate the lower-$J$ transitions when adopting the large $R_e$ found in the east-west direction, or underestimate them when adopting a smaller average $R_e$.


\begin{figure}[t]
  \centering
  \includegraphics[width=90mm]{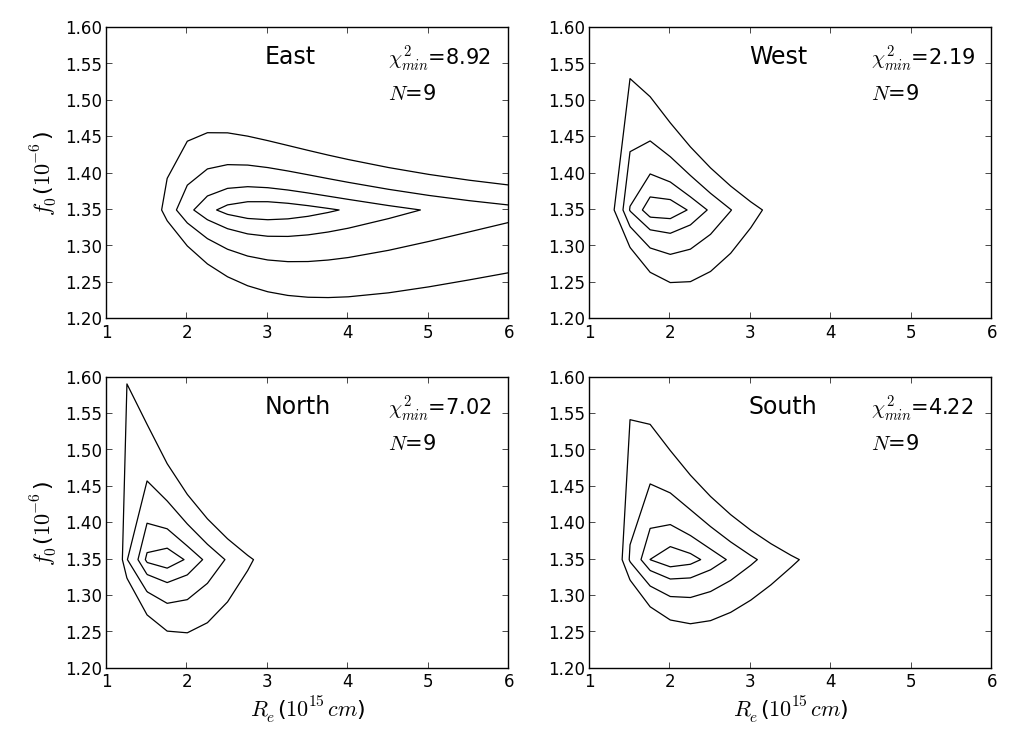}
  \caption[]{\label{chi2map2}
    Comparison between four $\chi^2_{red}$ maps derived using the intensities of H$^{13}$CN(4-3) at radial offset positions towards the east, west, north and south. The min $\chi^2_{red}$ values and the number of observational constraints that are used in the calculations are written in each panel.
  }
\label{chi2map2}  
\end{figure}

\subsection{Comparison with previous studies}\label{Comparison}

We compared our results with those of previous studies. 
The H$^{12}$CN circumstellar abundance value of $(5.0 \pm 2)\times 10^{-5}$ reported here for R Scl is higher than previously reported values of $8.1 \times 10^{-6}$ and $1.2 \times 10^{-5}$ by \cite{Olofsson93} and W04, respectively. 
It should be noted that \cite{Olofsson93} derived their abundance through a different method (see equation 3 in \cite{Olofsson93}).
However, our result is consistent with the median value of the H$^{12}$CN fractional abundances $f_0$ = 3.0 $\times 10^{-5}$ for a sample of 25 carbon stars studied by \cite{Schoier13} and the average value of $f_0$ = 4.6 $\times 10^{-5}$ for a sample of five carbon stars studied by \cite{Lindqvist00}. The derived H$^{12}$CN abundance here is also in good agreement with results of LTE stellar atmosphere models $f$=$(1-5) \times 10^{-5}$ reported for carbon-type stars by \citet{Willacy98, Cherchneff06}.


The $e$-folding radius of $R_e$ = (2.0 $\pm$ 0.25) $\times 10^{15}$ cm derived here is smaller than the values reported by W04 and \cite{Olofsson93} ($r = 1.5\times 10^{16}$ cm and $r = 3.6\times 10^{15}$ cm, respectively). 
However, the result reported in \cite{Olofsson93} is not based on solving the radiative transfer equation (see equation 6 in \cite{Olofsson93}).
The excitation analysis of W04 considered a fixed $f_{\rm HCN}$ throughout the envelope out to the outer radius $r = 1.5\times 10^{16}$ cm to fit the multi-transition spectra. However this radius is not consistent with the $J$=1-0 ATCA intensity map, which requires a smaller envelope of $ r= 5.5 \times 10^{15}$ cm and a higher photospheric abundance.

We speculate that the difference between the two envelope sizes derived based on the interferometric observations of H$^{12}$CN(1-0) by W04 ($\sim$ 360 AU) and H$^{13}$CN(4-3) here ($\sim$ 130 AU) could be due to the influence of the binary companion at $\sim$ 60 AU. As seen in CO in the outer envelope, the binary interaction causes a spiral density pattern that will alter the physical conditions in the outflow. The relative magnitude of the density pattern grows rapidly beyond the binary companion \citep[e.g.][]{Kim12} and the east-west HCN extension could be the observational indication of the start of the density pattern. The wave might have a stronger effect on the excitation of H$^{12}$CN(1-0) than that of H13CN(4-3).


\subsection{Comparison of the isotopologue ratios}\label{ratios comparison}

We have derived a ratio of circumstellar H$^{12}$CN/H$^{13}$CN = 26.3 $\pm$ 11.9 for R Scl which is consistent with the photospheric atomic carbon ratio of $^{12}$C/$^{13}$C 
$\sim$ 19 $\pm$ 6 reported by \cite{Lambert86}.
The derived isotopologue ratio is not affected by the limitation in the modelling (e.g. considering the effects of the binary companion on the physical condition), since the abundances are equally affected by these limitations.


The CN molecule is also photodissociated in the continuum \citep{Qadi13}. Hence, the $^{12}$CN/$^{13}$CN isotopologue ratio is expected to follow the $^{12}$C/$^{13}$C isotopic ratio. An intensity ratio of $^{12}$CN(1-0)/$^{13}$CN(1-0) $\sim$ 24 for R Scl reported by \cite{Olofsson96} is consistent with the photospheric $^{12}$C/$^{13}$C and the H$^{12}$CN/H$^{13}$CN ratio reported here. 

At the same time, the authors of V13 find an average value of $^{12}$CO/$^{13}$CO $\sim$19 in the detached shell, consistent with the atomic carbon photospheric estimates, whereas they derive a lower limit of $^{12}$CO/$^{13}$CO > 60 for the present-day mass loss.

Since H$^{12}$CN/H$^{13}$CN and $^{12}$CN/$^{13}$CN ratios are consistent with the $^{12}$C/$^{13}$C ratio, the most probable scenario of explaining the unexpectedly high $^{12}$CO/$^{13}$CO ratio in the inner CSE of R Scl is CO isotopologue selective-shielding.
The extra photodissociation of less abundant $^{13}$CO is most likely by the internal UV radiation as proposed by V13.

In addition, \cite{Olofsson96} have reported an intensity ratio of $^{12}$CN(1-0)/H$^{12}$CN(1-0) $\sim 4.3$ which is higher by a factor of two from the average value $2.0\pm0.7 $ reported for C-type stars. 
Assuming that HCN photodissociation is solely responsible for producing CN in the circumstellar environment, the ratio reported by \cite{Olofsson96} also supports the extra UV radiation in the CSE of R Scl.

Consequently, taking isotopologue self-shielding of molecules which are dissociated in discrete bands in the UV, e.g. $^{12}$CO/$^{13}$CO, into account is very important in the determination of the isotopologue ratios. Moreover, the isotopologue ratios of these molecules are not always a reliable tracer of the atomic carbon isotope ratio $^{12}$C/$^{13}$C in UV irradiated regions.
The discrepancy between the two ratios can be considerable in 
optically thick regions where the isotopologue self-shielding has more impact.

\section{Conclusion}\label{conclusion}

We have performed a detailed non-LTE excitation analysis of H$^{12}$CN and H$^{13}$CN in the CSE of R Scl. The derived H$^{12}$CN/H$^{13}$CN = 26.3 $\pm$ 11.9 is consistent with the photospheric ratio of $^{12}$C/$^{13}$C $\sim$ 19 $\pm$ 6 reported by \cite{Lambert86}. 

It is clearly shown that constraining the molecular distribution size, the fractional abundance and the radial expansion velocity profile in the CSE requires high-resolution spatially-resolved observations. 

Our results show that the circumstellar H$^{12}$CN/H$^{13}$CN ratio is a more reliable tracer of the photospheric $^{12}$C/$^{13}$C ratio than the circumstellar $^{12}$CO/$^{13}$CO ratio in the UV irradiated region. 
These results also support the isotopologue selective-shielding of CO as the reason for the lacking of $^{13}$CO in the inner CSE of R Scl as previously claimed by V13. 
The extra photodissociation of $^{13}$CO is most likely due to the internal radiation either from the binary companion or chromospheric activity.
This indicates the important role of internal UV radiation as well as the ISRF on the chemical composition of the CSEs. Thus, should be considered in the chemical-physical modelling of the CSEs around evolved stars.

In a more general context, the most abundant molecules in astrophysical regions that control the chemistry, for example H$_2$, CO, N$_2$, C$_2$H$_2$, have sharp discrete absorption bands in UV \citep{Lee84}, which leads to isotopic fractionation in the UV irradiated regions. Thus, considering the isotopologue selective-shielding is very important in astrochemical models of AGB envelopes and other irradiated environments.

We suggest that the comparison between the photospheric $^{12}$C/$^{13}$C ratio and the circumstellar $^{12}$CO/$^{13}$CO and H$^{12}$CN/H$^{13}$CN ratios can indirectly trace the internal embedded UV sources, which are difficult to observe directly, in the CSEs of evolved stars. 

This comparison cannot distinguish between the contributions from different potential UV sources such as ISRF, the chromospheric activity and active binary companions. This requires spatially-resolved observations of the photodissociated products such as atomic carbon.
 

\begin{acknowledgements}
This paper makes use of ALMA data from project No. ADS/ JAO.ALMA$\#$2012.1.00097.S. ALMA is a partnership of ESO (representing its member states), the NSF (USA) and NINS (Japan), together with the NRC (Canada), NSC and ASIAA (Taiwan) and KASI (Republic of Korea), in cooperation with the Republic of Chile. The Joint ALMA Observatory is operated by ESO, AUI/NRAO and NAOJ. This work is supported by ERC consolidator grant 614264. M.M. has received funding from the People Programme (Marie Curie Actions) of the EU’s FP7 (FP7/2007-2013) under REA grant agreement No. 623898.11.

\end{acknowledgements}




\bibliographystyle{aa} 
\bibliography{refrences} 


\begin{appendix} 

\section{}

\begin{figure}[hbtp]
  \centering
  \begin{minipage}[b]{0.48\textwidth}
    \includegraphics[width=\textwidth]{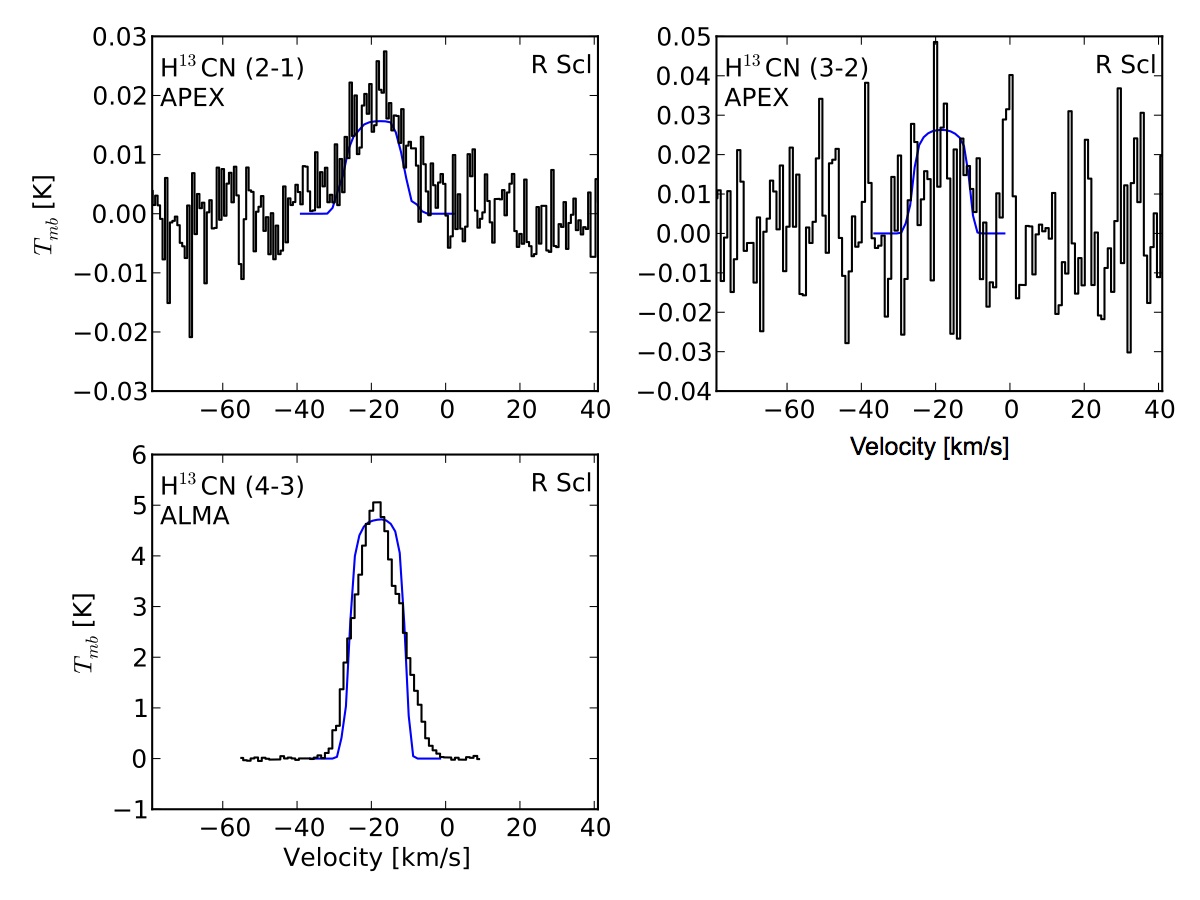}
    \caption{Line emission of H$^{13}$CN towards R Scl (black) overlaid with the model results (blue). Molecular transitions and the telescope used to get data are written in each panel.}
      \label{Toyh13cnSpectra}
  \end{minipage}
  \hfill
  \begin{minipage}[b]{0.48\textwidth}
    \includegraphics[width=\textwidth]{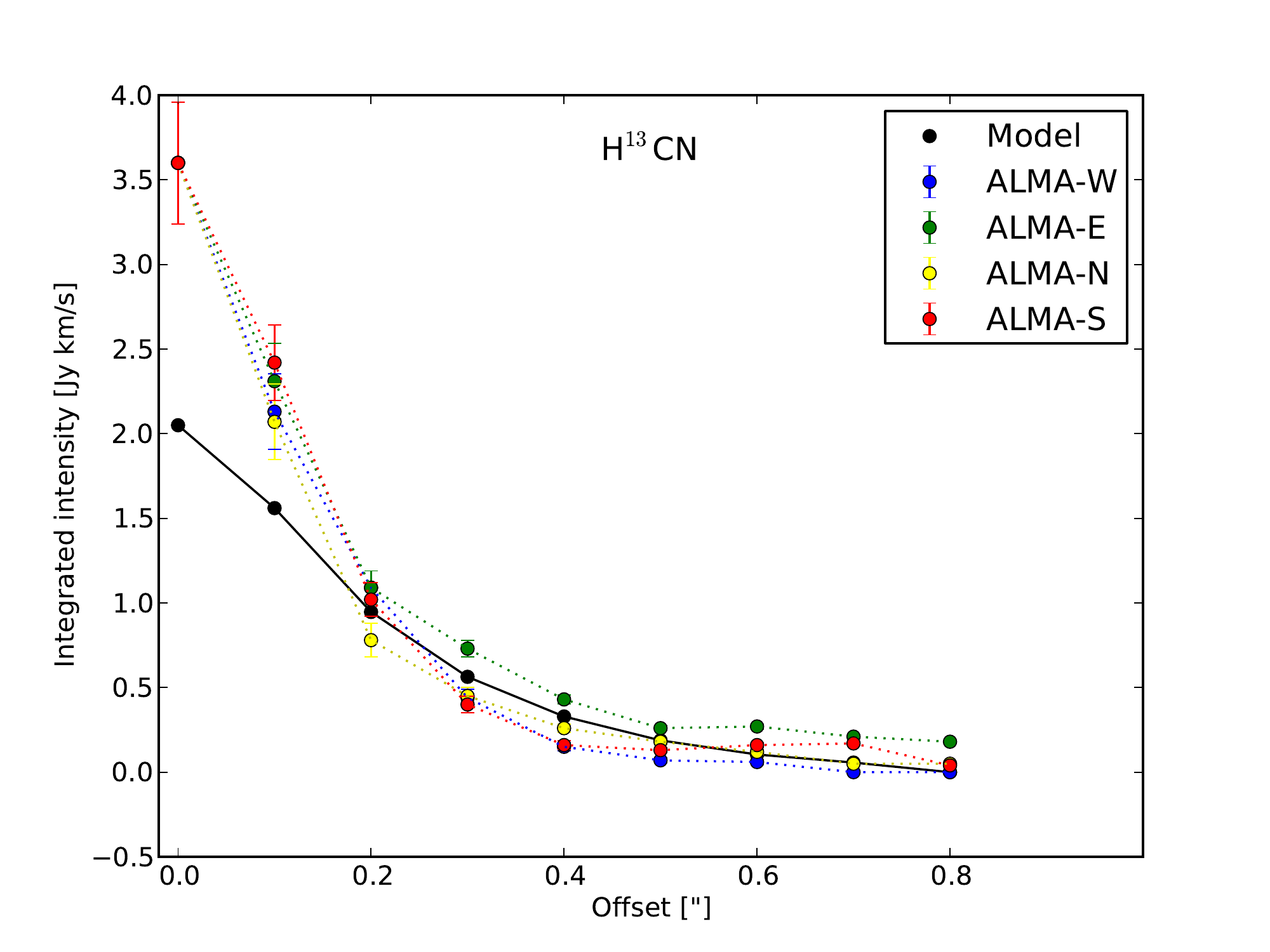}
    \caption{Comparison of the ALMA integrated intensities of H$^{13}$CN(4-3) at radial offset points in the CSE of R Scl towards the west, east, north and south from the centre of the star with the model results. Error-bars on the observational points show 10$\%$ uncertainty on the flux calibration.}
      \label{Toyh13cnRadial}
  \end{minipage}
\end{figure}

An example model which fits the total intensities of H$^{13}$CN(2-1, 4-3) spectra very well, but is not consistent with the integrated intensities at radial offset positions from ALMA observations,  Figs. \ref{Toyh13cnSpectra} and \ref{Toyh13cnRadial}.
This model has a radial expansion velocity profile with values b = 0.5 and $V_\infty = 8.5$. As it was mentioned in Sect. \ref{HCN-V}, the models with b < 2.5 give a double peak shape intensity profiles in the inner part ($r\leqslant0.2''$). Fig. \ref{ToyVexp} shows the intensity profile at the central point of the star from the model and ALMA observations. 
This model has a larger abundance of $f_0 = 8.0 \times 10^{-7}$ and a smaller radius of $R_e = 5.0 \times 10^{15}$ cm compared to our best model which was discussed in Sect. \ref{Results}. Thus, the best-fitting criteria based on the single-dish observations requires a smaller envelope and larger abundance.
This model shows the importance of using spatially resolved interferometric data in constraining the adjustable parameters in the modelling of the CSE.


\begin{figure}[t]
  \centering
  \includegraphics[width=75mm]{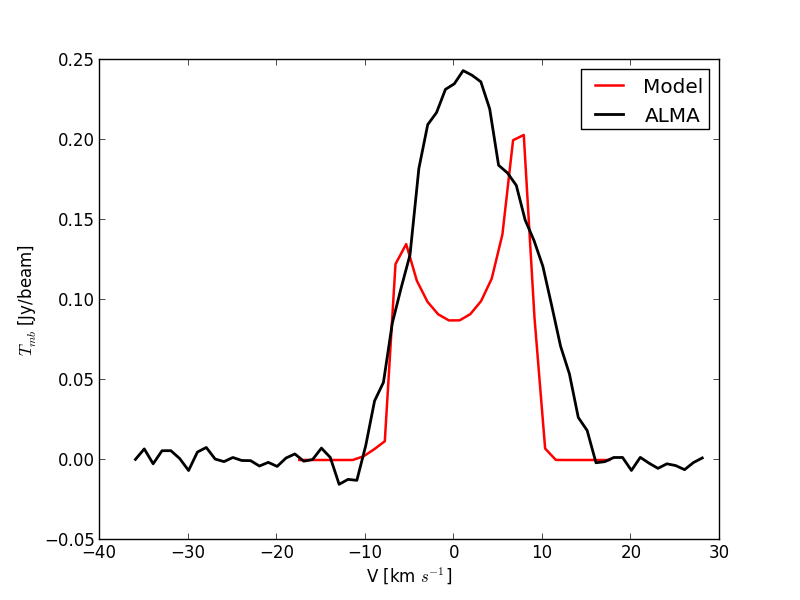}
  \caption[]{\label{ToyVexp}
      Comparison of the H$^{13}$CN(4-3) line emission towards R Scl at the centre of the star from ALMA observations and a model with fast accelerating wind (b = 0.5).}
\label{ToyVexp}  
\end{figure}

 \end{appendix}

\end{document}